 \documentclass[prl,twocolumn,showpacs,superscriptaddress,floatfix]{revtex4}
\usepackage{graphicx}%
\usepackage{dcolumn}%
\usepackage{bm}%
\usepackage{latexsym}
\usepackage{amsmath}

\def\be{\begin{equation}}
\def\ee{\end{equation}}
\def\bea{\begin{eqnarray}}
\def\eea{\end{eqnarray}}

\usepackage{natbib}

\begin{document}

\title{Real time cumulant approach for charge transfer satellites in x-ray
photoemission spectra}

\author{J. J. Kas}
\author{F. D. Vila}
\author{J. J. Rehr}

\affiliation{Dept. of Physics, Univ. of Washington, Seattle, WA 98195-1560}

\author{S. Chambers}
\affiliation{Physical Sciences Division, Pacific Northwest National Laboratory, Richland, WA 99352}

\date{\today}

\begin{abstract}
  X-ray photoemission spectra generally exhibit satellite features
in addition to the quasi-particle peaks due to many-body excitations,
which have been of considerable theoretical and experimental interest.
However, the satellites attributed to charge-transfer
(CT) excitations in correlated materials have proved difficult to
calculate from first
principles.  Here we report a real-time, real-space approach  for such
calculations based on a cumulant representation of the core-hole Green's
function and time-dependent density functional theory.
This approach also yields an interpretation of CT
satellites in terms of a complex oscillatory, transient response to a
suddenly created core hole.  Illustrative results for 
TiO$_2$ and NiO
are in good agreement with experiment.
\end{abstract}

\date{\today}
\maketitle
 Core-level x-ray photoemission spectra (XPS) often provides a direct probe of 
many-body excitations. For deep core levels and high energy photoelectrons,
the photocurrent $J_k(\omega)$ is roughly proportional to the 
core-hole spectral function $A_c(\omega)$,
and hence the excitations are reflected by satellite features in $A_c(\omega)$.
Thus theories of XPS beyond the quasi-particle approximation have 
been of considerable interest \cite{bergland64,caroli73,chang72,almbladh85,almbladh86,inglesfield81,inglesfield83,hmi,bard85,hedin99review}.
While there has been substantial
recent progress in \textit{ab initio} descriptions of
plasmon satellites in various materials \cite{aryasetiawan,guzzo,lischner}, first principles
  calculations of charge-transfer (CT) 
satellites have been elusive, especially in correlated materials
such as transition metal and actinide oxides.
These localized excitations have been attributed to the dynamic response
of a system to the creation of a deep core hole:
qualitatively the empty localized states, e.g., the transition metal
$d$-states,
of the photo-excited atom are pulled below the filled ligand levels, and
charge is transferred from the surrounding (ligand) atoms to screen
the core hole.
This process is reflected in the spectra as a well-screened state
of the core hole at higher energy,
and a poorly-screened satellite at lower energy (Fig.\ 1).
 Several 
approaches with various degrees of sophistication have been introduced
to treat this behavior.  Phenomenological models such as the single
impurity Anderson model , charge tranfer multiplet theory, and tight-binding
models have been used, with parameters derived from experiment \cite{degrootbook,lgh}.
First principles methods based
on configuration interaction techniques have also been applied to
CT excitations \cite{ikeno2011,bagus93,bagus2010}, but those
methods are computationally
intensive, and limited to a very small clusters of atoms. Thus none
of those approaches can be used to determine 
details such as spatial extent of these excitations. 

Recently cumulant expansion techniques have been found to explain
satellites in the XPS of weakly correlated systems due to
multiple-plasmon excitations \cite{aryasetiawan,guzzo,guzzo2012,lischner}, 
that are not captured by the conventional
GW approximation of Hedin \cite{hedingw}.
It is therefore of interest to investigate whether the cumulant approach can
be extended to treat the XPS for more correlated systems. This
approach is based on an exponential representation
of the core-hole Green's function $g_c(t)$ \cite{ND,langreth70},
and the spectral function is obtained from its imaginary part,
\begin{figure}
  \includegraphics[height=0.95\columnwidth, angle=-90]{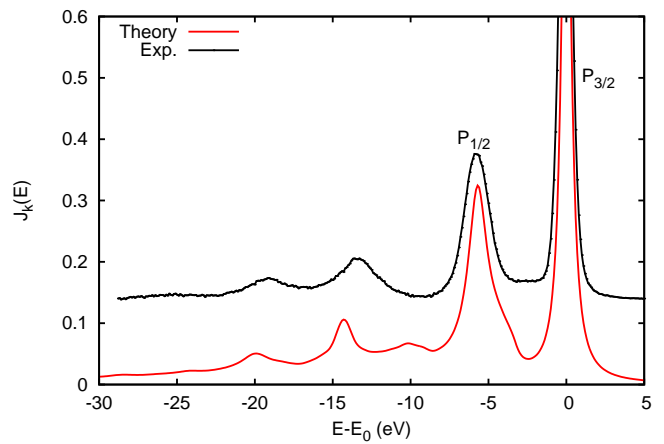}\vspace{3 mm}
  \caption{(color online). Comparison of the calculated XPS using the approach of this work
(red) and experimental (black) Ti 2p$_{3/2}$ and 2p$_{1/2}$ XPS of
TiO$_2$-rutile.  Each of the spin-orbit split quasiparticle peaks at
0 and -6 eV, exhibits a strong charge-transfer satellite at an
excitation energy $\omega_c \approx 14$ eV below.}
  \label{fig:RutileXPS}
\end{figure}
\begin{eqnarray}
  g_c(t) &=& g_c^{0}(t)e^{C(t)},\quad g_c^0=-\theta(-t)e^{-i\epsilon_c t} \\
  A_{c}(\omega) &=& -\frac{1}{\pi}{\rm Im}\int dt\, e^{i\omega t}g_c(t),
  \label{eq:spfcngf}
\end{eqnarray}
where $C(t)$ is the cumulant, and $\theta(t)$ is the unit step function. Throughout this paper we use atomic units
$e=\hbar=m=1$.
Following Langreth \cite{langreth70}, $C(t)$ can be
approximated to second order in the core-hole potential by 
\begin{equation}
\label{eq:cum}
  C(t) =  \sum_{\bm{q,q'}} V_{\bm{q}}^{*} V_{\bm{q'}} \int d\omega
S(\bm{q},\bm{q'},\omega)\,\frac{e^{i\omega t} -i \omega t -1}{\omega^{2}}.
\end{equation}
Here $V_{q}$ is the core-hole potential in momentum space, and
$S(\bm{q},\bm{q'},\omega)$ is the dynamic structure factor, which is related to
the density-density correlation function
\begin{equation}
  S(\bm{q}, \bm{q'},\omega) = \int \frac{dt}{2 \pi}e^{i \omega t}\langle \rho_{\bm{q}}(t)\rho_{\bm{q'}}(0)\rangle.
\end{equation}
Formally the cumulant expansion describes the transfer of spectral weight from
the quasi-particle peak to a series of satellites at frequencies $n\omega$
with an overall spectral function that preserves spectral weight.
This approach has its roots in the theory of Nozi\`{e}res and de Dominicis 
for edge singularities in core level x-ray spectra,
where the cumulant is derived from the linked-cluster theorem \cite{ND}.
For the case of a deep hole coupled to plasmons, the 
cumulant representation is exact \cite{langreth70}.
The time dependence $[\exp(i\omega t)-i\omega t -1]/ (\omega^{2})$
arises from the transient nature of the core-hole potential, which 
  turns on at time $t=0$, and off at time $t$. Considering this 
behavior and the localization of the core-hole, we are 
led to consider a real-space, real-time approach which is not limited
to very small clusters.  Real-time methods  can be advantageous
as they require little additional computational time beyond ground state DFT
calculations.  Here, we have adopted a real-time,
time-dependent density functional theory formalism (RT-TDDFT)
inspired by the work of Bertsch and Yabana \cite{yabana96}
for calculations of optical response.  This  formalism has since been
applied to both linear and non-linear optical response
in a variety of systems
\cite{yabana2006,takimoto2007,vila2010,otobe2009}. Recently this
this real-time method has been extended
to core level excitations \cite{ajlee}.
However, to our knowledge it has not previously been applied to
understand charge-transfer excitations.

\begin{figure}
  \includegraphics[height=\columnwidth, angle = -90]{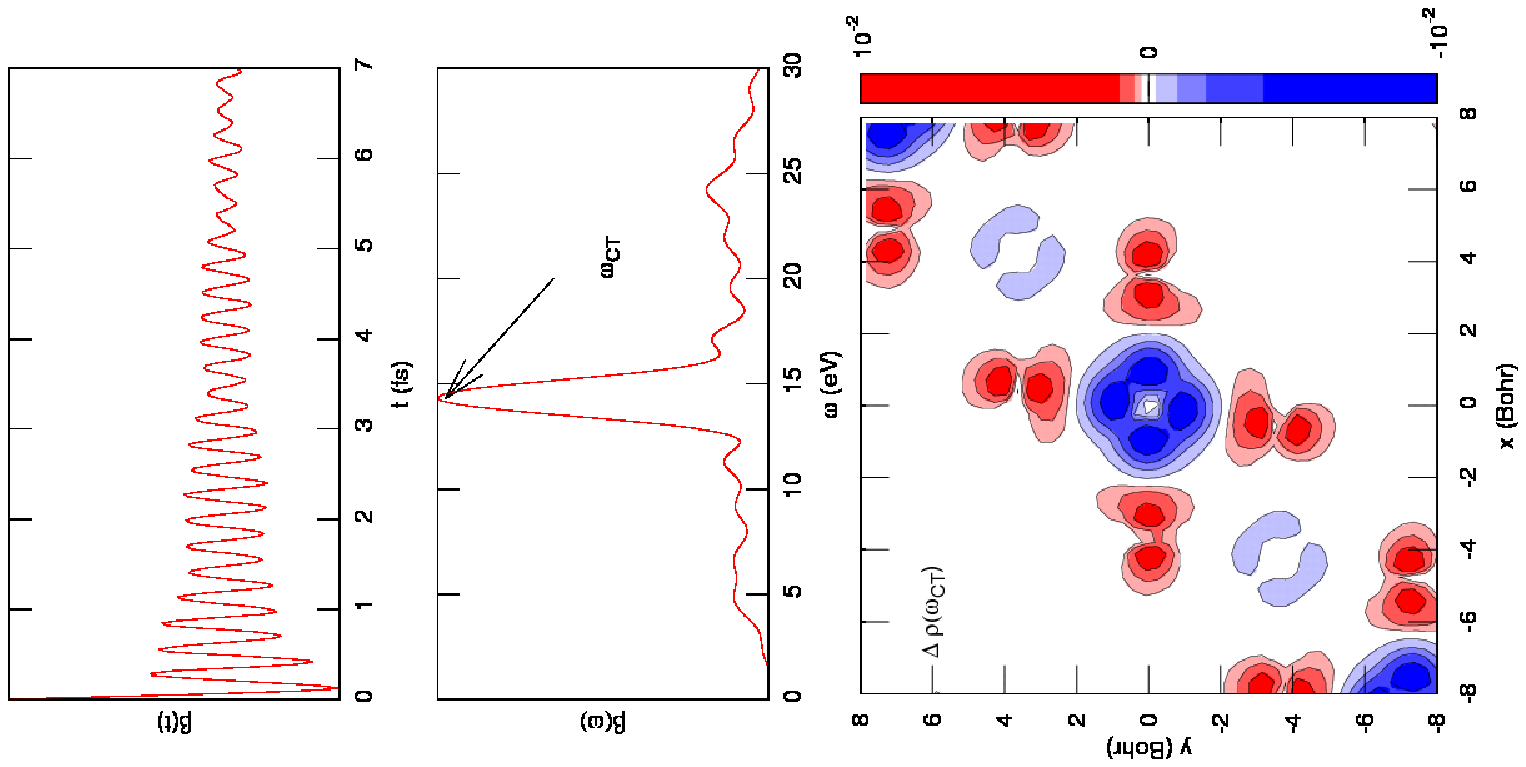}
  \caption{(color online). Top: Real time core-hole response function, $\beta(t)$ as a function of $t$. Note the transient response (fast relaxation of electrons) in the first fraction of a femtosecond. Middle: Core response function $\beta(\omega)$. Bottom: Excited state
density $\delta \rho(\omega_{c})$ at the ``charge transfer''
energy $\omega_{c}\approx14.3$ eV, denoted by the arrow in the middle plot.}
  \label{fig:denw}
\end{figure}

Our theory is briefly summarized below. We consider the
excitation of an electron in a deep core level $|c\rangle$ 
to an unoccupied photoelectron level with momentum $\bm {k}$
by a high energy x-ray.  The XPS photocurrent 
is given formally by the golden rule \cite{hedin99review},
\begin{equation}
  J_{k}(\omega) =  \sum_{s}\left|\langle N-1,s;
  \bm{k}|\Delta|N\rangle\right|^{2} ,
\end{equation}
where $|N\rangle$ is the $N$ electron ground state, $|N-1,s;\bm{k}$ is an
excited state characterized by photoelectron $\bm{k}$ and the $N-1$
electron system in state $s$,
and $\Delta$ is the dipole transition operator. If we
ignore all interactions between the photoelectron and the rest of the
system, the photocurrent can be expressed in terms of the one electron
spectral function $A_i(\omega)$ for a given level $i$,
\begin{equation}
  J_{k}(\omega) = \sum_{i}|\Delta_{ki}|^{2}A_{i}(\omega),
\end{equation}
where we have also assumed that the spectral function $A(\omega)$ is
diagonal in states $|i\rangle$. For deep core electrons and high
energy photoelectrons ($k \gg k_{F}$),
the dipole matrix elements are approximately
constant, and consequently the contribution to the photocurrent from
a given core level $c$ is proportional to the core-hole spectral function,
\begin{equation}
  J_{kc}(\omega) \propto A_{c}(\omega),
\end{equation}
where $A_c(\omega)$ is calculated from Eq.\ (\ref{eq:spfcngf}).
Transforming the equation for $C(t)$ to real-space
we obtain
\begin{eqnarray}
  C(t) &=& \int d^{3}r d^{3}r'\, d\omega\, V(\bm{r}) 
\delta\rho(\bm{r},\omega) \frac{e^{i\omega t}-i\omega t -1}{\omega^{2}}, \\
  \delta\rho(\omega) &=& \int d^{3}r S(\bm{r},\bm{r'};\omega) V(\bm{r'}).
\end{eqnarray}
The cumulant can therefore  be reexpressed as 
\begin{eqnarray}
  C(t) &=& \int d\omega \beta(\omega) \frac{e^{i\omega t}-i\omega t -1}
 {\omega^{2}}, \\
  \beta(\omega) &=& \int d^{3}r V(\bm{r})\delta\rho(\bm{r},\omega).
\end{eqnarray}
Here $\beta(\omega)$ is the excitation spectrum of the
\textit{effective} or ``quasi-bosons" \cite{hedin99review},
i.e., the  charge neutral excitations of the system.
This function is expected to exhibit peaks at the dominant excitation
frequencies, and
can be calculated in terms of the density fluctuations
$\delta\rho({\bf r},t)$.
The Fourier transform of $\beta(\omega)$ is given by
\begin{equation}
  \beta(t) = \frac{d^{2}C(t)}{dt^2} = \int d^{3}r V(\bm{r})\delta\rho(\bm{r},t).
\end{equation}
Physically $\beta(t)$ represents the potential fluctuations in the
response of the electrons
to an ``external'' time dependent perturbation that turns on at
time zero, i.e. $H^{(1)}(t) =
V(\bm{r})\theta(t)$. In contrast to optical spectra, however,
$\beta(t)$ is
dominated by mono-pole (i.e., $s$-like) response about the absorbing
atom. To illustrate this behavior the top and middle panels
of Fig.\ (\ref{fig:denw}) show this
response to a Ti core-hole in rutile TiO$_{2}$ in real-time (top), as
well as frequency space (middle). Although in principle, one might
calculate $\beta(\omega)$ directly from Eq.~(\ref{eq:cum}), using either TDDFT or BSE to obtain the dynamic structure factor, the localized nature of the
core-hole makes our real-space implementation very efficient for
CT excitations.

\begin{figure}[h!]
  \includegraphics[height=0.95\columnwidth, angle = -90]{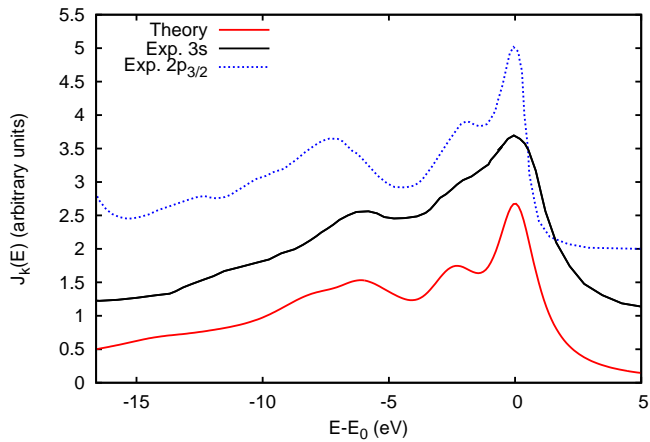}\vspace{3 mm}
  \caption{(color online) Calculated Ni 3s core-level spectrum for NiO (red) compared to experimental 3s (black) 2p$_{3/2}$ (blue dashed) XPS results \cite{altieri2000}.}
  \label{fig:NiOXPS}
\end{figure}

We have implemented this theory within a real-time TDDFT
extension \cite{takimoto2007} of the SIESTA code.
  The time-evolution is carried out using the Crank-Nicolson
propagator and an efficient basis of localized atomic
orbitals \cite{cranknicolson,siesta}. The detailed structure of the highly
localized core hole is not crucial, so we have simply modeled $V_c(r)$
as a Yukawa potential flattened inside a small radius to avoid
the singular behavior at $r=0$.
Thus this potential contains a single parameter that
characterizes its  strength, and thus the strength of the satellites in the 
spectral function. The response is then
calculated by relaxing the system to its ground state, turning on the
core-hole potential at time $t=0$, and then propagating the system to obtain the
induced time-dependent density fluctuations $\delta\rho({\bf r}, t)$. In order to uphold the relation between the second-order cumulant and the time-dependent density, we also scale the potential to stay withing the linear response regime, and rescale the resulting $\beta(t)$ accordingly.  
 Finally, the Green's function is formed according to
Eq.\ (\ref{eq:spfcngf}), and then Fourier transformed to obtain the spectral
function. The spin orbit splitting in TiO$_2$ is also treated as a parameter,
and it is assumed that the two excitations (p$_{1/2}$ and p$_{3/2}$) are
independent. An examination of the Ni 2p XPS of NiO (Fig.~\ref{fig:NiOXPS}) 
suggests that approximation may explain part of the discrepancy between
our calculations and experiment.

Fig.\ \ref{fig:RutileXPS} shows our calculated core-hole spectral
function (red) for Rutile TiO$_2$ compared to experimental XPS (black
crosses). The two largest peaks at $\approx 0$ and $-6$ eV
are the main quasi-particle peaks corresponding to the excitation of the
$p_{1/2}$ and $p_{3/2}$
(i.e., $L_2$ and $L_3$ edge) states respectively, split
by the 6 eV spin-orbit interaction.
Each of these main peaks has an associated CT satellite centered about
$14$ eV below, i.e., at about -14 and -20, respectively. These satellites
are qualitatively reproduced by the calculations, albeit with an
excitation energy that is slightly larger than that observed in
the experiment.

It is interesting to note that the response for TiO$_2$
is dominated by a fairly well defined frequency given by
the dominant charge-transfer excitation $\omega_c\approx 14$ eV 
(see top and middle plots of Fig.\ \ref{fig:denw}). This is not the case when the
core-hole is placed on the oxygen atom, indicating that these
excitations are strongly localized on the Ti atoms.
 Note also
the pronounced transient behavior in the first few femtoseconds,
and the sharp decrease at the onset within a fraction of a fs.
Physically, these features
correspond to relaxation of the valence electrons by charge transfer to the
ionized ligand atom,
followed by oscillation around the ground state in the presence
of the core-hole. The damping within the first few fs is
due to the diffusion of the excitation onto the surrounding atoms.
This effect requires the presence of an extended system and
would not likely be captured by highly localized models.

In order to interpret the source of these satellites spatially
we have plotted the Fourier transform of the induced density fluctuations
$\delta \rho({\bf r},\omega_c)$
evaluated at the charge transfer excitation energy $\omega_{c}$
(lower plot) in Fig.~(\ref{fig:denw}). For reference, the calculated response function
$\beta(t)$ (top), and its Fourier transform $\beta(\omega)$ (middle)
are shown.
The density is plotted for points ${\bf r}$ in  a plane through
the Ti atom and the four nearest oxygen ligands atoms,
with the Ti atom at the center. Note that the corners are near the edges
of the supercell and therefore simply reflect the density
near the core-excited Ti atom. The plot clearly illustrates an oscillatory
transfer of electrons from the Ti atom to the ligands during the CT
excitations process. In addition, the shape of the density fluctuations
suggests a transfer of electrons from Ti $3d-$orbitals to
O $2p$-orbitals.

To illustrate a wider applicability, we have also carried out calculations for
NiO.
Fig.\ (\ref{fig:NiOXPS}) shows the experimental (blue) Ni 3s XPS of
NiO compared to our calculated results (red). Again the main peak (at 0 eV)
and largest satellite at approximately $-6$ eV are in qualitative
agreement with experiment.
The theory also reproduces a peak seen in the experiment at about
$2.1$ eV. 
For reference, the experimental Ni 2p$_{3/2}$ XPS is shown
as well. The higher energy resolution of this spectrum allows a
more detailed analysis, although there are clearly other differences
between the two experimental spectra. In particular,
it is interesting to note the difference in the energy and strength of the second major
satellite, signifying a role of either the shape of the core hole, or
a difference in the core-valence exchange interaction between the two
cases.  
We have also compared with previous calculations
of the NiO spectral function based on the non-orthogonal configuration interaction (NOCI) method \cite{hozoi2006}.
Table (\ref{tbl:energies}) shows the energies of the first three experimentally visible satellites relative to the energy of the main peak as calculated
with our RT-TDDFT method compared to those of Ref.\ \cite{hozoi2006}
and experiment.
\begin{table}
  \label{tbl:energies}
  \caption{Relative energy of the first three satellites in the spectrum as calculated with RT-TDDFT and NOCI, as well as those extracted from experimental data.}
 \begin{ruledtabular}
   \begin{tabular}{cccc}
     RT-TDDFT & NOCI & Exp. \\
     2.3 & 2.0 & 2.2 \\
     6.1 & 7.7 & 6.1 \\
     8.3 & 8.1 & 10.2
   \end{tabular}
 \end{ruledtabular}
\end{table}
The agreement between our RT-TDDFT calculations and the experimental values for the first two major excitations is reasonably good, while that for the third peak is too small by several eV, similar to the results of the NOCI calculations.

In conclusion we have developed a real-space, real-time formulation of
the core-hole spectral function based on the cumulant expansion
and TDDFT calculations of the cumulant.
The method is implemented using the RT-TDDFT extension of SIESTA,
and has been applied to calculations of CT excitations in
Rutile TiO$_2$ and NiO. The relative energies and amplitudes of
satellite peaks are in
semi-quantitative agreement with
experiment for for both oxides. In addition we have shown that the
excitations can be interpreted by inspection of the response in real
space and real time to the sudden appearance of a core hole. The response
is characterized by several time scales, in particular 
a transient
response within a fraction of a fs, corresponding to the
relaxation of the valence electrons to a new ground state, followed by
oscillatory charge transfer between the core and the ligand orbitals.
The CT excitations correspond to the transfer from localized
Ti states of $3d$ character to states of O $p$-character. While this
method is very promising in its own rite, especially for calculations
of deep-core XPS, these calculations could also be used to extract
parameters for models of x-ray absorption spectroscopies such as
charge transfer multiplet models, or the model of Lee, Gunnarson, and
Hedin \cite{lgh,calandra2012,klevak2014}, which includes both intrinsic and
extrinsic interactions as well as the interference between
them. Future plans include development of a more realistic core hole,
and extensions to include multiplet effects, e.g., in L edge spectra.


Acknowledgments - We thank G. Bertsch, L. Reining and A. Lee for
useful comments.
This work was supported by DOE Grant DE-FG03-97ER45623
(JJR and JJK) and DOE BES DMSE award number 10122 (SAC),
and was facilitated by the DOE Computational Materials
Science Network. One of us (JJR) also thanks the
the Kavli Institute for Theoretical Physics at UCSB, and 
the Laboratories des Solides Irrad´ıes at the ´Ecole Polytechnique,
Palaiseau for hospitality when parts of this work were carried out.


\bibliography{references}
\end{document}